# Chemical kinetic theory of aging

Alexey Kondyurin
Ewingar Scientific. Ewingar, NSW, Australia
kond@mailcity.com
ORCID: 0000-0002-8837-9734

**Abstract**

Theory of aging based on the kinetics of chemical reactions and the rules of natural selection of organisms is proposed. Biochemical processes in organism can be described in terms of chemical reaction kinetics. The evolutionary process of organisms is determined by the goal of continuing life, and the natural selection forced organisms to develop in an optimized way for survival and reproduction, after which any further development of the organisms did not matter for natural selection and therefore was not optimized. Accordingly, the ongoing biochemical processes in the organism after giving birth were not stabilised and continued casual, which led to imbalances and biochemical processes that did not contribute to health in the organism, resulting in aging and death.
Artificial balancing of these processes can potentially stabilize of the kinetics of biochemical reactions and the life can continues almost indefinitely.

**Introduction**

For today there are a number of theories of human aging (Jin, 2010). All of these theories can be divided into two large groups. The first group consists of theories of programmed death, and the second group consists of theories of violations or errors.

Programming theories claim that aging is programmed and follows a biological schedule. This schedule can be implemented as a sequence of turning genes on or off. Aging is observed as an age deficit (Davidovic et al., 2010). It has been shown in a number of experiments that there is a "biological clock" in the cell that limits the cell's ability to continuously proliferate. Based on these experiments, the limit of human life is estimated to no more than 100-120 years (Gerhard & Cristofalo, 1992; Hayflick, 1996). But other experiments have shown that a cell can proliferate indefinitely. Experiments on limited cell proliferation and cell death rather refer to an unnatural cell proliferation environment.

Another theory of programmed aging is based on the organism's use of hormones to regulate the aging process (van Heemst, 2010). Indeed, all organs of the organism are regulated by hormones that are secreted by the organs of the endocrine system. Together, the neurologic and endocrine systems determine many processes of the organism's metabolic activity. One of the functions of hormone system is the reproductive activity of the organism. The concentration of hormones responsible for the reproductive function of the organism is directly related to the active period of women and men to produce a child. Other hormones regulate biological rhythms and are responsible for activating or suppressing the immune system, and a number of hormones are responsible for nutrient absorption. Changes in the neurologic and endocrine systems occur with age concentration of hormones change in the organism. However, the artificial intake of additional doses of hormones does not help to increase life time, although in some cases it can temporarily solve complications of aging in an organism, in particular, reproductive function. It also remains unclear why the concentration of hormones changes during life time.

Another view on the problem of aging is associated with the immune system. The activity of the immune system decreases with time and it ceases to protect the organism, which leads to aging and death (Cornelius, 1972; Rozemuller et al., 2005). Indeed, the decrease of the immune system activity leads to a decrease in the organism's resistance to diseases and an increase in the risk of cancer. This is shown in a number of cases, for example, with HIV patients and patients with implanted organs where the immune system was supressed. However, the mechanism by which the activity of the immune system declines in the elderly remains unclear. On the other hand, methods of increasing the activity of the immune system does not lead to an increase in life time.

The theory of aging based on the shortening of telomerase can be attributed to the same group. Over time, during cell division, telomerase is shortened, the cell stops proliferating, and this causes the death of the whole organism (Flores et al., 2005; Hayflick & Moorhead, 1961; Herbig et al., 2006). However, artificial methods of telomerase lengthening also does not lead to an increase in the lifespan of the whole organism.

Theories of wear and tear of organisms are based on how everyday wear and tear affects the human organism's ability to maintain itself. For example, the theory of the standard of living is based on the basic oxygen metabolism, which if accelerated causes a shorter lifespan of an organism (Hulbert et al., 2007). The cells and organs of the organism then wear out and stop functioning, and the organ tissues can no longer be renewed and the organism dies. This theory intersects with theories of programmed death. However, it should be noted that physical exercise can increase organism functionality, but does not shorten or prolong life. As a variant of this theory, it is assumed that the processes of growth and maintenance compete for bodily resources (Rollo, 2010).

One of the ways for the accumulation of errors in organism functionality is the theory of protein cross-links, which causes disturbances in biochemical processes in the organism (Bjorksten, 1968; Bjorksten & Tenhu, 1990). The cross-linking of collagen protein is mainly considered, which hinder the transport of nutrients and reduce tissue elasticity. The result can be seen by comparing the soft, supple skin of a child with the hard, wrinkled skin of an older person. The chemical reaction of crosslinks can be accelerated by unsaturated fats, metal ions, and by radiation. The use of a special diet that excludes the intake of such cross-linking agents can reduce the rate of these cross-linking process. Using antioxidants as inhibitors of cross-links that have been caused by radiation damage also helps, however, the problem of aging and death has not yet been solved by diet and vitamin intake.

The theory of the accumulation of defects such as free radicals, including ones in cell DNA, belongs to the same class of these theories (Gerschman et al., 1954; Harman, 1956; Afanas'ev, 2010). Replication errors can occur in all stages of DNA and RNA replication. Accumulation of errors above a certain level can lead to improper synthesis of cell proteins and death. However, accumulation of significant amounts of DNA and protein synthesis errors in old cells – which could cause disruption of cell and organism functionality – has not been experimentally detected (Hayflick, 1996; Goldstein, 1993; Schneider, 1992). This theory is also contradicted by data on people who have received large doses of radiation during catastrophes and yet maintained a long heathy life, despite the extremely high value of the number of damages to DNA and other active molecules in the organism.

A similar theory is based on free radical accumulation and damage in cell membranes in the mitochondria, lysosomes, and nuclear membranes, which leads to lipid oxidation, and so then the transport of substances into cells and the functionality of the cell are impaired. The organism has its own mechanisms to protect and resist the oxidation processes, however, over time they become insufficient. Animal experiments have shown that the administration of antioxidants reduced the occurrence of heart disease and prevented the occurrence of cancer. Even though the intake of antioxidants slows down the degeneration of the nervous and immunity system, the intake of antioxidants does not give a significant increase in life expectancy. In addition, it is unclear why the organism's ability to fight lipid oxidation declines over time.

There are also a number of theories of aging based on social and psychological factors associated with higher nervous activity. These theories are not considered here because they do not relate to organisms without higher nervous activity. At the same time, aging processes for such animals also exist and are similar to the aging processes within the human organism.

Thus, so far none of these theories fully explains the aging process and do not show ways to slow it down, or stop it (Davidovic et al., 2010). More importantly, we still do not understand why we age, what is the aging process, how did it appear, and what for?

**Hypothesis**

To understand the aging process of an organism, we must first consider the origin and development of life. Modern theories of the origin of life on Earth are based on chemical reactions in the synthesis of organic molecules. The synthesis of these complex organic molecules from a mixture of ammonia, carbon dioxide and other molecules dissolved in water is possible, as shown in a number of experiments. However, such organic molecules can both form and decompose under the influence of environmental factors. We should assume that at some point some of these special organic molecules appeared. The key characteristic of such molecules as the precursors of life is the ability to exactly replicate themselves. That is, such molecules should be able to collect the necessary molecules from their environment to clone themselves. The clone production is a chemical process, that can be described by chemical reaction kinetics rules.

RNA molecules have this ability under certain conditions, however, this does not mean that exactly those RNA molecules that we observe now appeared in the proto-soup. Perhaps these were the simplified forms of precursors of to today's known RNA molecules, which still had the main property of precise replication. There are only descendants of these self-replicating molecules, making them our ancestors. The rest of the organic molecules that did not have the ability to clone themselves formed in the proto-soup eventually broke up and disappeared, so now there is little information left about them.

The further development of such molecules entailed the possibility of protein synthesis that helps the synthesis of its clones. The protein synthesis and functionalisation are a chemical process following rules of chemical reaction kinetics. Then, a lipid membrane from the external environment was synthesised, and the cells with the membrane appeared to be more protected against harsh environment. As a result, the protocell separated from environment and could replicate a clone appeared. Further development of the protocell led to more complex mechanisms of protection from the external environment and obtaining the necessary substances to maintain this protection. The chemical processes and their kinetics parameters were optimised to get a stable cell. But the main property in such cells - the ability to replication - has been preserved.

Later, cells began to unite in conglomerates, where cells found ways of communication and regulation. The ability of cells to distribute the responsibilities of functions improves the preservation of these cells as part of a conglomerate in the environment. At each stage of development, the cell conglomerate become more complex and probability of successful reproduction increased. It was achieved through the complication of biochemical processes. This process was determined by the rules of natural selection. The optimization of biochemical processes in the cell conglomerate leaded to survive most consistently compared to other cell conglomerates. And the cells' ability of replication remained as a key-factor.

Then, such conglomerates formed multicellular organisms, where different cells in the organism would have a certain functionality. The specialization of cells within a multicellular organism allows for efficient division of functionality and a greater potential for adaptation and survival in changing environments. Thus, the main parameter of the survival of the organism's species is the ability to replicate itself. The main evolution driving force is the rules of natural selection. The main mechanism of the organism evolution is changing of biochemical reactions and its kinetic parameters. These are the main factors that drive further genetic development improvement.

And how are these processes related to the aging process?

For a unicellular organism, the aging as such does not exist. Unicellular organisms are able to divide endlessly. There is no ageing mechanism for a single cell. Even a cell separately taken out of an animal or humans can divide endlessly in certain conditions with nutrients, gas, medium, temperature, and sterility as it is proven in the experiments. This has been proven by evolution of life on Earth, which has a history of more than 3 billion years.

But at some stage of the evolutionary process of the cell conglomerates, a problem arose of the difficulty of maintaining simultaneous endless replication and protection of a conglomerate of cells from the external environment. During this process a new path of biochemical reactions spontaneously solved this problem leading to the successful protection of the conglomerate up until it could provide successful offspring. But the fact that after replication with this new path, whether such a conglomerate could exist became unimportant. After replication, the conglomerate degraded and nothing could stop its degradation, since there was no optimisation of biochemical reactions to maintain such a conglomerate. Any processes that were linked to survival after replication were no longer deemed important by the rules of natural selection. The offspring of the conglomerate had already been produced and the replication repeated. The structure of RNA (and later DNA) was preserved in nature. The former conglomerate had been removed from evolution and became waste. At this stage, the infinite lifecycle of the cell conglomerate was sacrificed to obtain the most optimal parameters for the survival of the cell conglomerate in the external environment and the production of offspring.

From this point of view, the process of aging and death is seen as optional for multicellular organisms, but it happened as a result of evolution. So far, no one knows whether it is possible to slow down the aging process, or if the aging process is strictly related to the organism's ability to develop and produce offspring, and if the exclusion of one property will cause the exclusion of another property.

On the other hand, starting from the replication of the first RNA molecules, the kinetics of these transformations can be described in terms of a chemical reaction. Further complication of the structure of replicating RNA molecules, and later of cells does not change the laws of chemical transformation. Therefore, the biochemical transformations in the cell and organism must obey the laws of the kinetics of a chemical reaction. This means that the concentration of a certain substance B must be determined by the chemical kinetic equations with the reaction constants for obtaining such a substance from substance A, and the kinetic equations for the chemical reactions of the transformation of substance B into substance C with appropriate coefficients depending on the reaction conditions, including the presence of catalysts. Of course, no one can simplify the real kinetics of biochemical reactions in a both a cell or an organism to a few equations, but there is no reason to assert that the principles of chemical kinetics in a cell and an organism will differ significantly from the principles of chemical reaction kinetics. This simplification can help to understand the principles of aging as a biochemical process determined by the concentration of certain substances in the cells and organism.

**Model**

Let us consider a model of the kinetics of chemical reactions. Suppose there is a substance we will call "A" in a certain concentration in a certain volume. This substance can be converted as a result of a chemical reaction into substance we will call "B" at a certain rate. Suppose the rate of this reaction is determined solely by the concentration of substance A. Then the differential reaction equation will correspond to the first-order reaction equation:

$$\frac{\partial [A]}{\partial t} = -k_1[A] \tag{1}$$

where $k_1$ is the reaction rate constant. This equation describes the most chemical reactions.

Let us assume that substance B can also be transformed into a substance we will call "C". Let say the reaction kinetics also correspond to the first order of the reaction with the reaction constant $k_2$. Then the equation for the substance B is following:

$$\frac{\partial [B]}{\partial t} = k_1[A] - k_2[B] \tag{2}$$

Next, consider the chemical reactions with substances C, D, E, F, and so on. For ease of consideration, let us assume that the reactions correspond to first-order reactions. The system of the equations would be the following:

$$\frac{\partial [A]}{\partial t} = -k_1[A]$$

$$\frac{\partial [B]}{\partial t} = k_1[A] - k_2[B]$$

$$\frac{\partial [C]}{\partial t} = k_2[B] - k_3[C] \tag{3}$$

$$\frac{\partial [D]}{\partial t} = k_3[C] - k_4[D]$$

...

$$\frac{\partial [X_n]}{\partial t} = k_{n-1}[X_{n-1}] - k_n[X_n]$$

Let's limit the calculation to substance I, which we will install last in the chain of reactions. The solution of equations can be represented in the form of a graph (Fig. 1).

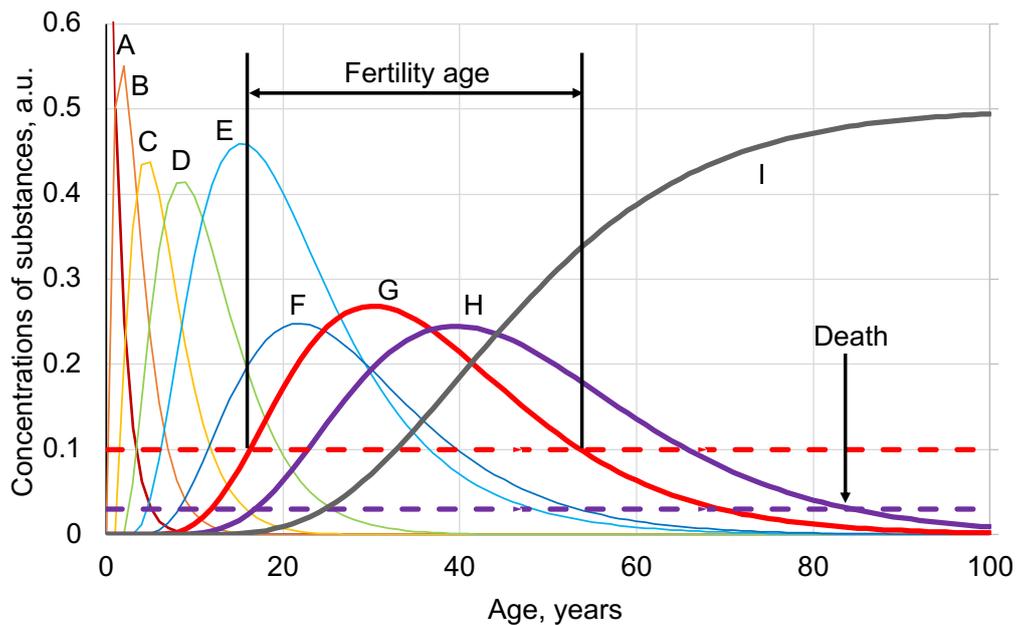

Fig.1. Kinetics of chemical reactions according to the system of equations (3). Explanation in the text.

Assume that these reactions occur in the human organism from birth or conceiving. If we assume that substance G is a regulator of the hormones responsible for reproductive functions, like a regulator of testosterone and estradiol for example, then the ability of a person to produce offspring appears at a certain age in youth and disappears at old age. As the concentration increases, in order to produce offspring, it is necessary that the hormonal level be above a certain limit, which is shown in the graph by the dotted line. Thus, the ability to reproduce offspring appears in a person some time after his birth and is determined by a chain of chemical reactions, and disappears due to subsequent chemical reactions. For example, in this model calculation, the age of fertility starts at 17 years and ends at 53 years.

Let's assume that substance H is responsible for the vital functions of the organism in old age. Then lowering this substance below a certain limit will cause the death of a person. A decrease in the concentration of this substance H can also be caused by chemical reactions ongoing in the organism. In this calculation, death occurs at the age of 84 years

The organism at the end of fertility is withdrawn from the process of evolution. If older than this age, the organism does not produce offspring and does not participate in the process of natural selection. Accordingly, further development of the organism is not influenced by natural selection. Due to the fact that the running chemical processes in the organism proceed further, after the end of fertility, according to the theory of the kinetics of chemical reactions, it makes no sense to expect that these processes are aimed the maintaining organism functionality. Chemical reactions in the organism occurring after the end of fertility may unbalance the vital processes in the organism and as a result the organism dies.

Let's consider some cases. Let us assume that the reaction constant of the transformation of substance F is 30% less than in the previous case (Fig.1), while keeping all other parameters same. Then the fertile age begins at 20 years and lasts until 60 years (Fig.2). Death occurs at the age of 96 years.

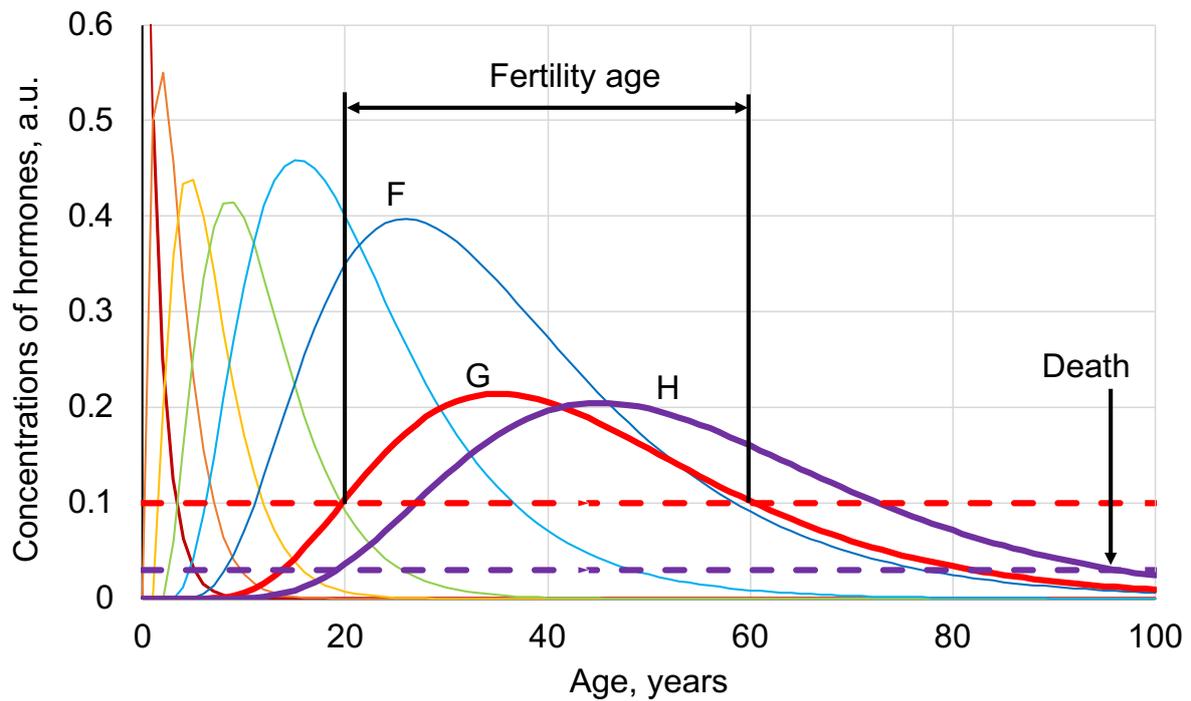

Fig.2. Kinetics of chemical reactions according to the system of equations (3). The reaction rate of the substance F is reduced by 30% in comparison with the calculation results in Fig.1. Explanation in the text.

Consider a situation where a person gets an injection of substance F periodically starting from the age of 50 every 10 years. Then the solution of the equations (3) will look like this (Fig.3).

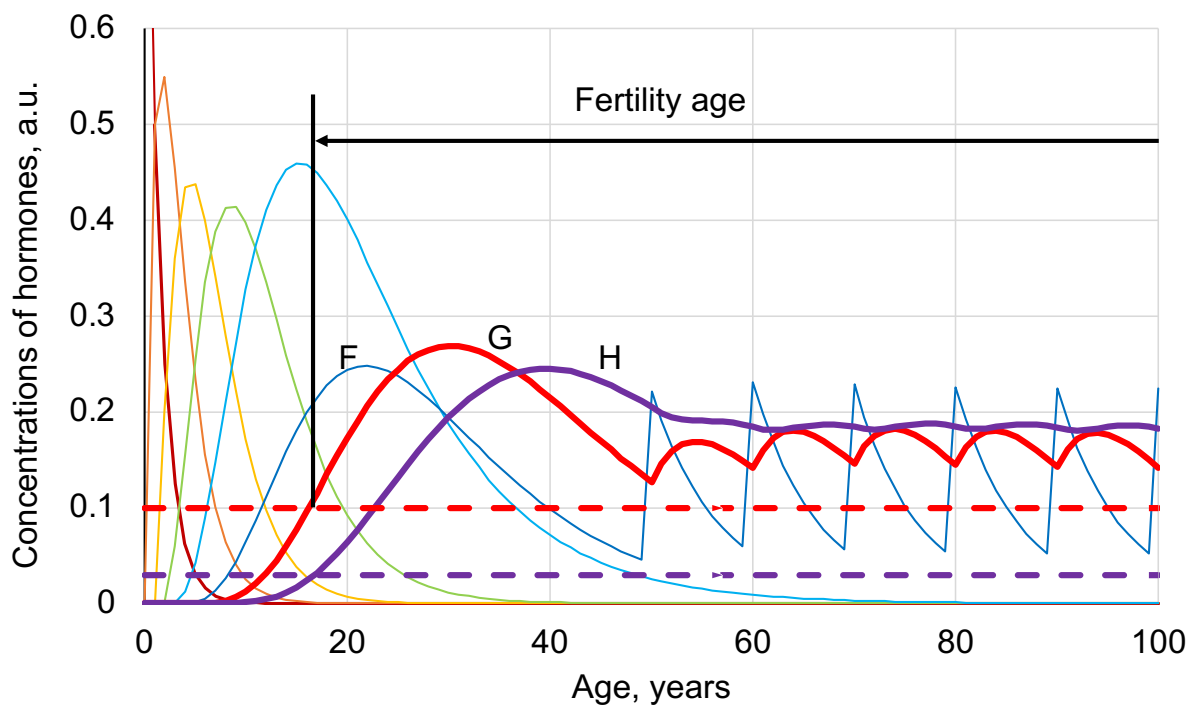

Fig.3. Kinetics of chemical reactions according to the system of equations (3). In comparison with the previous solution, the substance F is added into the organism every 10 years. Explanation in the text.

The introduction of substance F changes the course of the chemical reactions so that substance G and H will never cross the limits of fertility loss or death. Accordingly, the organism will not age and will always be able to produce offspring. Of course, this does not solve the problem of diseases, catastrophes and other events and their impact on human life. But the aging process according to this model can be stopped.

**Discussion**

What is known today about the concentration of biochemical regulators of vital activity in relation to such a model of chemical kinetics in the organism? Unfortunately, there is practically no monitoring of biochemically important regulatory compounds in the organism of the same healthy person from birth to death. The most complete data on the dynamics of the concentration of biochemically important compounds in the organism can be found related to hormones.

It is known now that hormones are regulators of an organ functionality in the organism. They cannot represent components of chain biochemical reactions that determine the development of the organism. However, the production and release of hormones during the life of the organism is regulated by certain agents. A release of the hormones and their concentration are provided by the regulators of the endocrine system, which manages the growth and development of the whole organism. Therefore, hormones can be considered as indicators of the activity and concentration of regulators of biochemical processes in the organism. Some examples of well-known measurements of hormone concentrations in a healthy organism in life time can be presented.

One example is Somatotropin as a growth hormone in mice (Fig.4a). This growth hormone is responsible for bone growth and organ development, but is also a hormone with distinct catabolic and anabolic functions in many tissue types (Vorotnikova et al., 2011; Huang et al., 2019). The concentration of the hormone falls within 8 weeks after the birth, when an animal becomes adult and fertile. Similar results can be found for hormone concentrations in other animals.

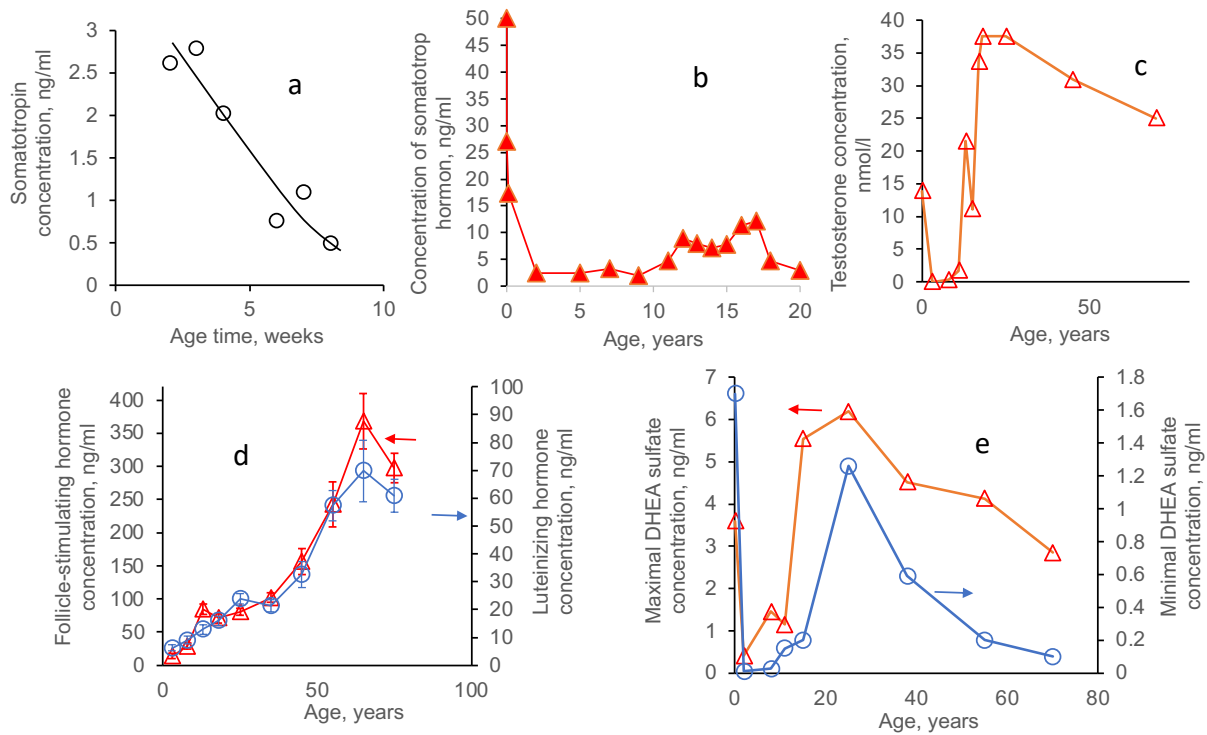

Fig.4. (a) Concentration of somatotropin growth hormone in mice with age by (Katelnikova, 2020). (b) Concentration of the somatotropin growth hormone in humans (boys, upper level) with age. First point is umbilical blood by (Smartlab, 2023; Nazarenko & Kishkun, 2006). (c) Top concentrations of Testosterone in men with age by (Travison et al., 2017; Sperling, 2014; Labcorp, 2021; Morgentaler, 2017; Morgentaler et al., 2014; Wang et al., 2008; Bhasin, 2011). (d) Concentration of Follicle-stimulating hormone and Luteinizing hormone in human with age by (Medbe, 2023). (e) Upper and bottom concentration of ehydroepiandrosterone sulfate (DHEA sulfate) in men with age by (Nazarenko & Kishkun, 2006).

The dependence of the concentration of somatotropin in human blood soon after birth shows similar trend (Fig.4b). The highest concentration of this hormone is observed in blood from the umbilical vein at birth. Furthermore, the concentration falls and remains constant until teenage years. During teenage years, the concentration of this hormone increases. Then, after the growth period in adolescence, the concentration of the growth hormone falls.

The concentration of the well-known hormone testosterone is high in the womb and falls shortly after birth (Fig.4c). In adolescence, during maturation, the concentration of testosterone rises sharply and reaches a maximum at the age of 20-25 years. Further, the concentration of testosterone continuously declines in old age. This hormone is responsible for reproductive functions, for the growth and development of bone and muscle tissue, as well as mood.

The concentration of follicle stimulating hormone (FSH) and luteinizing hormone (LH) increases throughout a person's life (Fig.4d). These hormones regulate the human reproductive system. A certain concentration of these, and other hormones is required for male and female fertility. Deviations in the concentration of these hormones from the optimal values (to lower or higher values) leads to infertility.

The concentration of the hormone DHEAS rises during adolescence, reaches a maximum around 18-25 years of age, and then falls (Fig.4e). This hormone belongs to neurosteroids, and is associated with memory. There is evidence that this hormone is also responsible for diseases of the cardiovascular system, diabetes, Parkinson's, Alzheimer's, and others.

The presented curves show that the concentrations of hormones change with age and the curves of changes have a characteristic shape for the curves of the concentrations of substances in chain reactions. Curves with sharp changes tend to be at a young age, curves with a maximum tend to be at a fertile age, and monotonically decreasing and monotonically increasing curves of hormone concentration tend to be observed in old ages. Basically, sharp changes in concentration are observed in childhood and at the age of maturation. Changes in old ages are more gradual. Most of the curves show falling hormone concentrations in old ages after the age of fertility.

Note that none of the above theories of aging provides an explanation of the behaviour of the concentrations of hormones and other biologically important substances in an organism. On the other hand, changes in the concentrations of biologically important regulatory substances in the organism seem to correlate with phases of change in organism functionality, including the reproductive phase and the old age phase. This confirms the correlation between the organism's functionality and the concentration of biologically important regulators of the vital activity of the organism. Of course, the real scheme of the biochemical reactions in the organism is much more complicate, but this very simple model demonstrates the main trend of the organism ageing.

This present model of aging based on the kinetics of chemical reactions answers to the fundamental questions posed in (Hayflick, 1996; Nazarenko & Kishkun, 2006):

1. Why do organisms undergo a progressive decline in physiological functions in the last part of their lives?
2. Why does the rate of aging vary within species and between species?
3. Why do some restrictions (reducing the calorie content of food, for example) slow down aging and lengthen life?

1. According to the general conclusions of the theory of the kinetics of chemical processes in a closed system, the concentrations of reactants decrease with the reaction time and the concentration of the final reaction products increase. Apparently, the same applies to the regulators of vital processes in the human organism and other living organisms. The initial concentration of these substances is set at conception or at the birth of an organism. Then a decrease in the concentration of regulators leads to a decrease in critically important hormones and, as a result, to a decrease in the physiological functions of the organism. Model shows this process at the last part of the life.

2. The course of a chemical reaction depends on the rate constant of the reaction. The difference in the rate constants of chemical reactions in the organism, apparently, is determined by the genome of the organism and may differ between organisms of the same and different species. The evidence is the difference in the absolute values of the concentrations of hormones and other biologically important compounds in the organisms of different individuals within the same and different species. The difference in the genome within the same species is small. The same small difference is observed in the duration of life within the same species. The difference in the genome of different species can be significant. There is also a significant difference in life expectancy between species. The model shows how a decrease in the rate constant of only one of the regulators by 30% shifts the age of fertility and the date of death of the organism.

3. An artificial change in the metabolic rate (for example, due to the calorie content of food) apparently also affects the rate of chemical reactions of the regulators responsible for the development of the organism. In the presented model, an example with a decrease in the rate constant of the reaction of a substance F lengthened the life of the organism.

These results should not be taken directly as a reason for an injection of some hormone into an organism. The organism is a rather complex system and the introduction of one or more hormones will create an imbalance in the organism, and can disrupt its normal functionality. It is necessary to find the key substances and/or regulators of the vital activity of the organism, the artificial regulation of which will change the dynamics of hormone concentrations. Currently, the accumulated results on the studies of the human organism are not enough to make a decision on the introduction of any drug or hormone to increase life expectancy. However, at present, there are no detailed studies of the kinetics of reactions in the organism of a healthy person (or animal) during his/her entire life, starting from birth or even from the moment of conception. Without the knowledge of the kinetics of the organism's biochemical reactions, the search for ways to regulate or stop the aging process is a random process. A knowledge of the kinetics of the regulators in the organism, knowledge of the kinetics of biochemical processes, including the patterns and chains of reactions of the regulators of the vital activity of the organism, such as cytokines, hormones, and others is required.

Thus, the theory of organism aging based on the kinetics of chemical reactions of developmental regulators can explain the aging dynamics in principle. Further detailed study of the kinetics of biochemical processes with age, starting from the moment of conception, may make it possible to regulate biochemical processes for life extension or even immortality.

**Conclusion**

Thus, it is assumed that the aging of an organism is not a planned and obligatory result of life, but is rather the result of one path of genealogical development of a multicellular organism, which was evolutionarily chosen by chance. If so, then there is a possibility that it might be possible to adjust the biochemical reactions of the organism so that the aging process be slowed down or even stopped.

**Statements and Declarations**

Author declares no financial support and no conflict of interests.